\documentclass[preprint]{aastex631}

\usepackage{amsmath}
\usepackage{amssymb}

\begin{document}

\title{Intermittent Electron-Only Reconnection at Lunar Mini-Magnetospheres}

\author[0000-0002-0786-7307]{A. Stanier}
\affiliation{Los Alamos National Laboratory, Los Alamos, New Mexico, USA}
\email{stanier@lanl.gov}

\author[0000-0002-0786-7307]{Li-Jen Chen}
\affiliation{NASA Goddard Space Flight Center, Greenbelt, MD, USA}

\author[0000-0002-0786-7307]{A. Le}
\affiliation{Los Alamos National Laboratory, Los Alamos, New Mexico, USA}

\author[0000-0002-0786-7307]{J. Halekas}
\affiliation{Department of Physics and Astronomy, University of Iowa, Iowa City, IA, USA}

\author[0000-0002-0786-7307]{R. Sawyer}
\affiliation{Department of Physics and Astronomy, University of Iowa, Iowa City, IA, USA}

\begin{abstract}

Lunar crustal magnetic anomalies (LCMA) are sub-ion-gyroradius structures that have been shown to stand off the solar wind (SW) plasma from the Moon's surface, forming shock-like discontinuities and reflecting incident SW protons. In this letter, the results of high-resolution, two-dimensional fully kinetic simulations show a bursty electron-only magnetic reconnection in the SW-LCMA interaction region, characterized by the quasi-periodic formation and ejection of magnetic islands and strong parallel electron flows along the X-point separator lines. The islands are observed to modify the magnetic pressure pile-up and Hall electric field above the LCMA, leading to sharp increases in reflected protons that drive electromagnetic fluctuations downstream and short distances upstream in the SW. 
\end{abstract}

 
\section{Introduction} \label{sec:intro}

The Moon does not have a global-scale magnetic field dipole to shield it from charged particles in the solar wind (SW). To lowest order, the SW particles are incident on and absorbed by the lunar surface to leave a deep void in the downstream wake~\citep{lyon1967explorer,colburn1967}. The interplanetary magnetic field (IMF) passes through the Moon with only a weak enhancement in the downstream wake~\citep{colburn1967}, that is supported by diamagnetic currents at the wake boundaries~\citep{owen1996,fatemi2013}.

However, this lowest order description is enriched by the presence lunar crustal magnetic anomalies (LCMAs), which are distributed in a complex and multi-polar pattern over the Moon's surface~\citep{coleman1972satellite,purucker2010global}. The scale size of these anomalies varies between $1 - 1000$ km~\citep{dyal1974} with surface magnetic field strengths of up to a few hundred $nT$, falling to a few $nT$ at $100$ km altitude.

Despite their small scale, LCMA can have an significant impact on the incident solar wind. Spacecraft measurements have detected reflection of up to 50\% of incident solar wind protons above LCMA~\citep{futaana2003,lue2011}, some of which can be energized by the convective electric field and even gyrate into the deepest part of the near-Moon wake~\citep{nishino2009,xu2020reflected}. This strong modulation of the solar wind flux at the LCMA can affect the surface albedo and cause the bright lunar swirl patterns visible on the lunar regolith~\citep{hood1980,deca2018reiner}. Further, sharp enhancements of the magnetic field have been measured downstream of the LCMAs close to the limb -- so-called ``limb-compressions"~\citep{russell1975compressions, lin1998,halekas2017}, that are suggestive of a shock-like structures forming upstream of the LCMA~\citep{lin1998,halekas2014}.

The SW-LCMA interaction has been investigated by numerical simulation using a variety of different plasma models. Early efforts used a single-fluid magnetohydrodynamics (MHD) description~\citep{harnett2003}. With the size of the interaction region being less than the ion skin depth, $d_i = c/\omega_{pi}$ with $c$ the speed of light and $\omega_{pi}$ the ion plasma frequency, this motivated studies using Hall-MHD~\citep{xie2015hall}, hybrid kinetic-ion fluid-electron~\citep{giacalone2015hybrid,fatemi2014hybrid}, and fully kinetic~\citep{deca2014prl,deca2015,bamford2016} models. 

An important question regarding SW-LCMA interaction concerns the possible occurrence and effects of magnetic reconnection~\citep{ji2022natrevphys} between the IMF and the LCMA fields. Earlier kinetic simulations noted the lack of typical reconnection signatures~\citep{deca2014prl,deca2015} such as bi-directional reconnection outflow jets. However, recent data from ARTEMIS at $15$ km altitude~\citep{sawyer2023} have measured the Hall electric field at LCMA, as well as the presence of solar wind electrons on closed magnetic field-lines -- an indication of reconnection having occurred between the IMF and the LCMA fields. Reconnection has also been observed at Mars~\citep{eastwood2008mars,halekas2009mars} -- another body without a global dipole field -- with the MAVEN spacecraft detecting reconnection occuring above a Martian crustal magnetic anomaly~\citep{harada2018mars}. 
If reconnection can indeed occur between the solar wind IMF and the LCMA, the sub-ion scale-size of the interaction region suggests that demagnetized ions would not form the characteristic bipolar reconnection outflow jets, and the regime may be more similar to electron-only reconnection observed in the Earth's magnetosheath~\citep{phan2018electron}. 

In this paper, we report 2D kinetic particle-in-cell simulation results that show a \textit{bursty} electron-only magnetic reconnection occurring in the SW-LCMA interaction region, characterized by the quasi-periodic formation and ejection of magnetic islands and parallel electron flows along the separator lines of the magnetic X-points. These islands were found to form only for simulations with close to realistic values of the proton-electron mass ratio, such that the electrons are strongly magnetized and the aspect ratio of the current layer is large enough to become tearing unstable. The presence of these islands is found to modify the pressure balance between the LCMA and the solar wind inflow, increasing both the magnetic field pile-up strength and the Hall electric field. The resulting quasi-periodic increases in the reflected ion fraction produce observable ``spike" features in the simulated ion density profiles, as well as rapidly varying magnetic field enhancements that can propagate both downstream and a short distance upstream  of the LCMA. 

\section{Numerical Simulations} \label{sec:setup}
In this paper, results are presented from 2D electromagnetic particle-in-cell simulations of the solar wind-LCMA interaction. The simulations are performed using the Vector Particle-In-Cell (VPIC) code~\citep{bowers08,bird2021vpic}.

\subsection{Simulation set-up and boundary conditions}

The solar wind plasma and interplanetary magnetic field are injected at a constant rate and orientation from the $x=0$ inflow boundary. The solar wind velocity vector is in the $+x$ direction, $\boldsymbol{v}_{SW} = v_{SW}\boldsymbol{\hat{x}}$, and the IMF lies in the $x-y$ plane making an angle $\theta_{IMF}=\arctan{\left(\frac{B_{y,IMF}}{B_{x,IMF}}\right)}$ from the solar wind direction. 

For the inflow boundary at $x=0$, and the top/bottom boundaries of the simulation domain (at $y=0, L_y$), we choose boundary conditions based upon the initial solar wind/IMF values. Particles colliding with these domain boundaries are absorbed, and new particles are sourced at the boundary at each step based upon the drifting Maxwellian distribution function used for the solar wind. The VPIC code uses a typical Yee-mesh~\citep{yee1966,birdsall2004} scheme for spatial discretization of the electromagnetic fields that is staggered in space, such that only the tangential electric field and normal magnetic field components lie exactly on the domain boundaries. The components are fixed equal to their initial values as $B_n = \boldsymbol{\hat{n}}\cdot \boldsymbol{B}_{IMF}(t=0)$ and $E_t = -v_{SW} \boldsymbol{\hat{x}}\times \boldsymbol{B}_{IMF}(t=0)$. (It was found that setting additional field components such as $B_t$ gave rise to unphysical fluctuations in the inflowing plasma). The spatial domain was sufficiently large (see below) that the locally strong LCMA fields become negligible at the domain boundaries, and the SW and IMF parameters there match the initial and boundary values used. 

All particles incident on the Moon and the outflow boundary at $x=L_x$ are absorbed, with their accumulated surface charge $\rho_{\textrm{surface}}$ being used to correct $\boldsymbol{\nabla} \cdot \boldsymbol{E} = \left(\rho_{\textrm{surface}} + \rho_{\textrm{plasma}}\right)/\epsilon_0$, and no particles (e.g. photoelectrons) are presently sourced from the Moon's surface. The IMF is allowed to pass freely through the Moon and the outflow boundary ($x=L_x$).

\subsection{\label{sec:params}Solar wind and LCMA parameters}
Solar wind plasma parameters and IMF strength are chosen based upon ARTEMIS spacecraft observations~\citep{halekas2014} of shock-like compression features being detected at high altitude (periselene $~420$ km) above the Lunar surface. The density of the solar wind $n_{SW} = 12.5$ cm$^{-3}$, the solar wind velocity $v_{SW} = 300$ km sec$^{-1}$, and the proton and electron temperatures are $T_p = T_e = 7$ eV. The IMF strength is $|B_{IMF}| = 7.5$ nT, where the orientation angle $\theta_{IMF} = - 30^\circ$. In terms of dimensionless parameters, these give an Alfv\'en Mach number $M_A = 6.7$, and solar wind plasma beta $\beta_i = \beta_e = 0.6$. The mass ratio is $m_i/m_e = 1836$. The ratio of the electron plasma to cyclotron frequencies is set to the value of $\omega_{pe}/\Omega_{ce} = 2$ in the results presented below, but we have checked the sensitivity of the results up to $\omega_{pe}/\Omega_{ce}=8$ (at lower mass ratio) without finding significant qualitative differences in the results.

The 2D LCMA fields are applied via line-dipole currents below the surface. For a 3D (ring) magnetic dipole with strength $3\times 10^{13}Am^{-2}$ this would stand-off the solar wind at a distance of $D_p = 35 \textrm{km} = 0.55 d_i$ from the dipole center, using the solar wind parameters specified above, and have a strength of $2$ nT at 100 km altitude. Here, for our 2D model, we determine the line dipole strength by matching this pressure stand-off distance to the 3D value. The line-dipole is buried $15$ km $= 0.23 d_i$ below the surface of the Moon, such that the effective pressure stand-off distance would be $20$ km $(0.32 d_i)$ above the surface. The LCMA is at a solar zenith angle of $-45^\circ$.

The spatial domain used is $12.5\times 12.5 d_i$ ($536\times 536 d_e$ with $1d_e \approx 1.5$ km). This is resolved with $8192 \times 8192$ cells to give a constant mesh size of $\Delta x = 0.065 d_e = 0.24 \lambda_D$, where $d_e$ is the electron skin depth and $\lambda_D$ is the electron Debye length. There are an average of $50$ particles per cell for each species. The simulation is run a significant time prior to the presented results ($t\Omega_{ci} >  8$), so that the solar wind passes through the domain and carries away any initial transients associated with the initial conditions. 

\section{Results} \label{sec:results}

\begin{figure}
    \centering
    \includegraphics[width=0.9\textwidth]{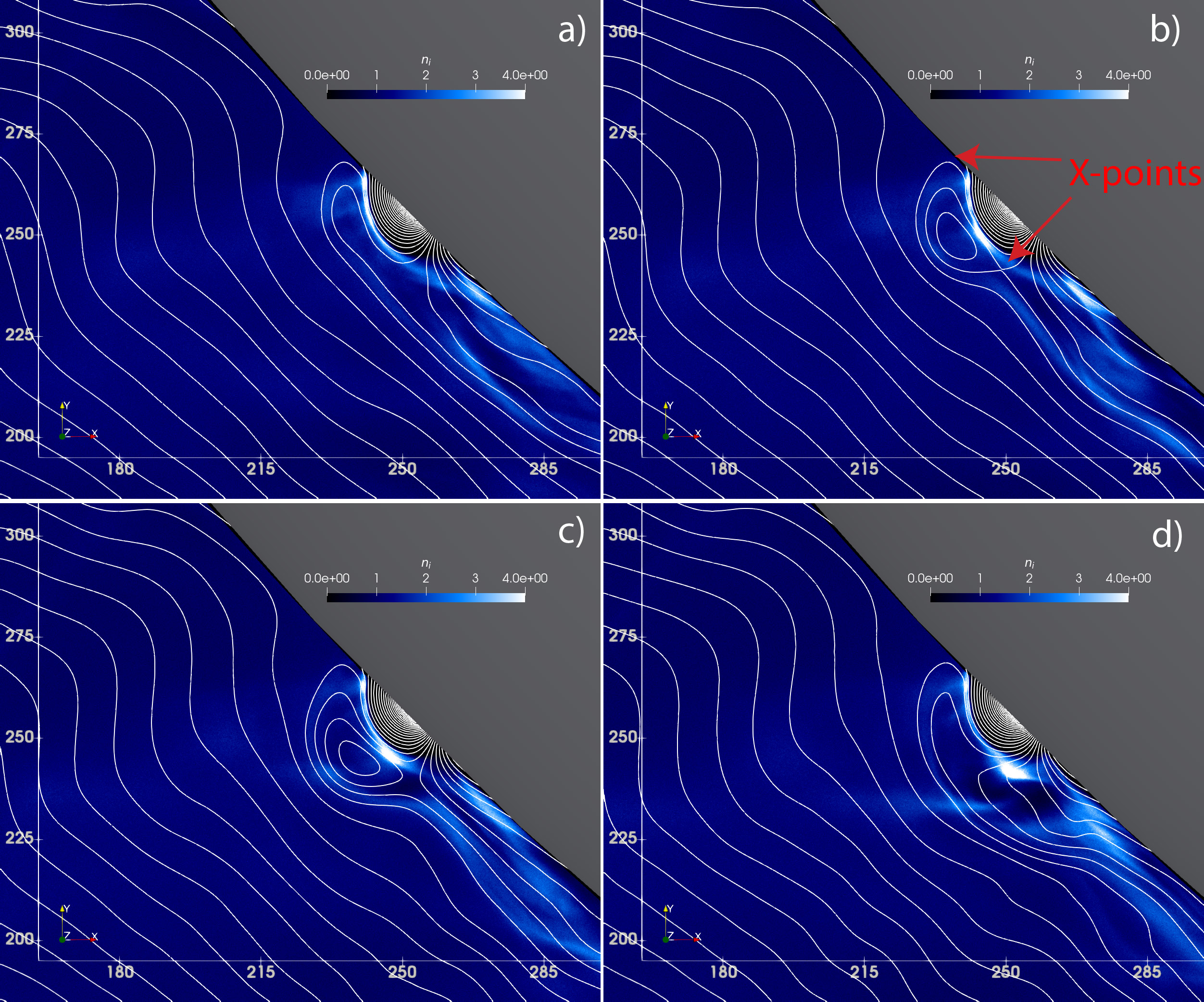}
    \caption{Ion density (color scale) normalized by the solar wind density $n_{\textrm{SW}}$, and magnetic field lines (white), shown at four snapshots separated by $0.05 \Omega_{ci}^{-1} = 92\Omega_{ce}^{-1}$. The Moon (gray) is at the upper right corner of each plot. The spatial coordinates are given in electron skin depths ($1 d_e \approx 1.5$ km). The region shown is only a small part of the full domain.}

    \label{fig:miniFTE-formation}
\end{figure}

Figure~\ref{fig:miniFTE-formation} shows the formation and ejection of a magnetic island, occuring on electron time-scale in the SW-LCMA interaction region. In the initial plot (1a), the X-point is located above the LCMA dipole-center line and close to the surface of the Moon. This X-point position is displaced upwards from the vacuum field location, indicative of large currents in the SW-LCMA interaction region, and this displacement is found to be noticeably larger for simulations with $m_i/m_e \gtrsim 400$. A tearing instability, with an associated electron stagnation point flow (see below), leads to the formation of a magnetic island (1b), which grows in size due to the reconnection of magnetic flux at the upper and lower X-points (1c). At its largest size the magnetic island is elongated with semi-major radius $25 d_e$ and semi-minor axis $12.5 d_e$. The magnetic island is then ejected downwards and reconnects with the open magnetic field-lines coming from the southern pole of the LCMA. These field-lines are strongly perturbed from their initial orientation parallel to the Moon's surface (1d), to generate a small-scale disturbance that propagates downstream above the Moon's surface (see below). The total time taken for the formation and ejection of the magnetic island is $\approx 0.2 \Omega_{ci}^{-1} \approx 367 \Omega_{ce}^{-1}$ based upon the interplanetary magnetic field strength $B_{\textrm{IMF}}$. 

\begin{figure}
    \centering
    \includegraphics[width=0.95\textwidth]{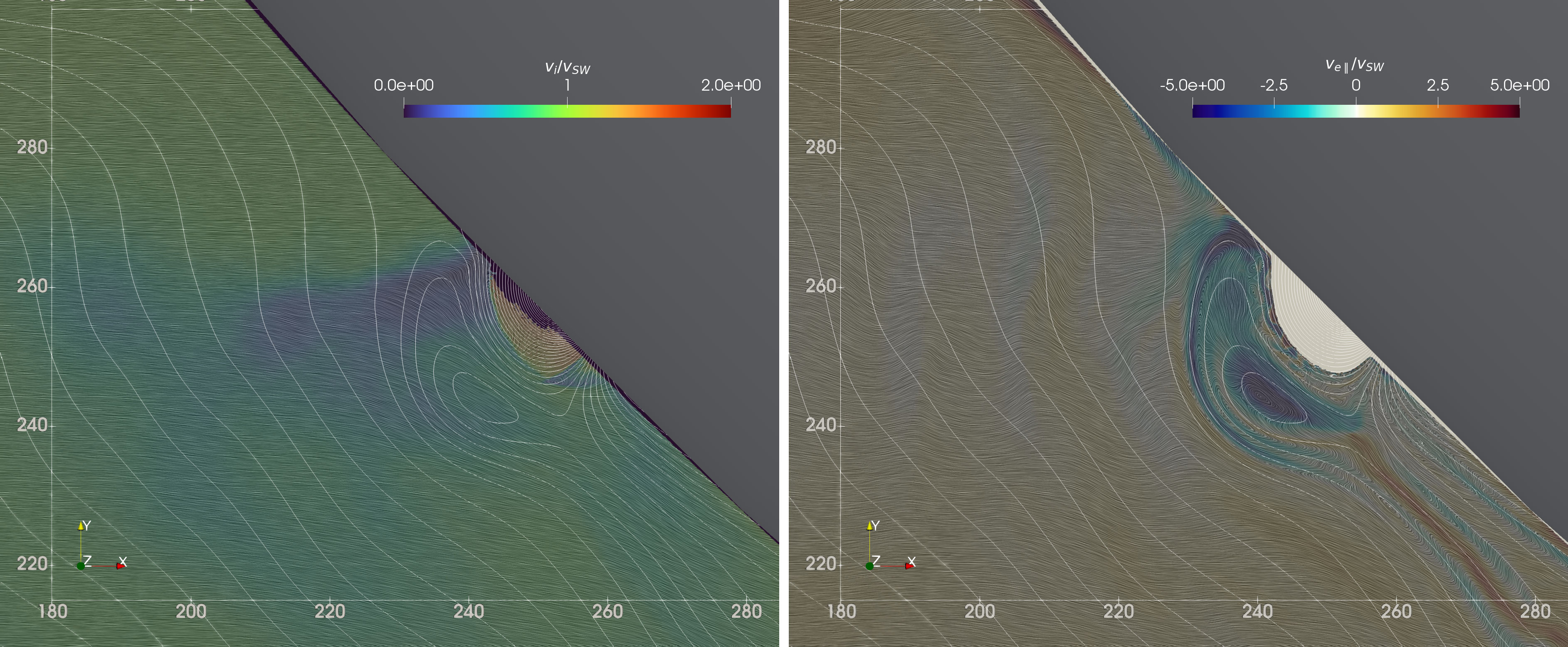}
    \caption{Line integral convolution (LIC) of the ion bulk velocity field (left) and the electron bulk velocity field (right). The ion plot is colored by the ratio of the ion to solar wind speed, and the electron plot is colored by the ratio of the electron parallel velocity to the solar wind speed. The white contours show the magnetic field lines and the Moon (gray) is in the upper right corner. The spatial coordinates are given in electron skin depth ($1 d_e \approx 1.5$ km).}
    \label{fig:streamlines}
\end{figure}

Figure~\ref{fig:streamlines} shows line integral convolution (LIC) plots of the ion and electron bulk velocities around the SW-LCMA interaction region in the presence of a magnetic island. Due to the small scale size of the LCMA compared with the ion kinetic scales, the bulk ion flows are only weakly deflected by the LCMA fields and magnetic island, although they are decelerated to a minimum velocity of $u_x \approx 0.2 v_{SW}$ along a line that passes through the vertical center of the dipole source. In particular, there are no discernible ion outflow jets associated with reconnection occurring at the X-points. In contrast, the electron flows are strongly modified by the magnetic topology. There are strong parallel flows of electrons along the topological separators of both of the X-points that reach about $3 v_{SW}$ at times when there is an island present.

\begin{figure}
    \centering
    \includegraphics[width=0.98\textwidth]{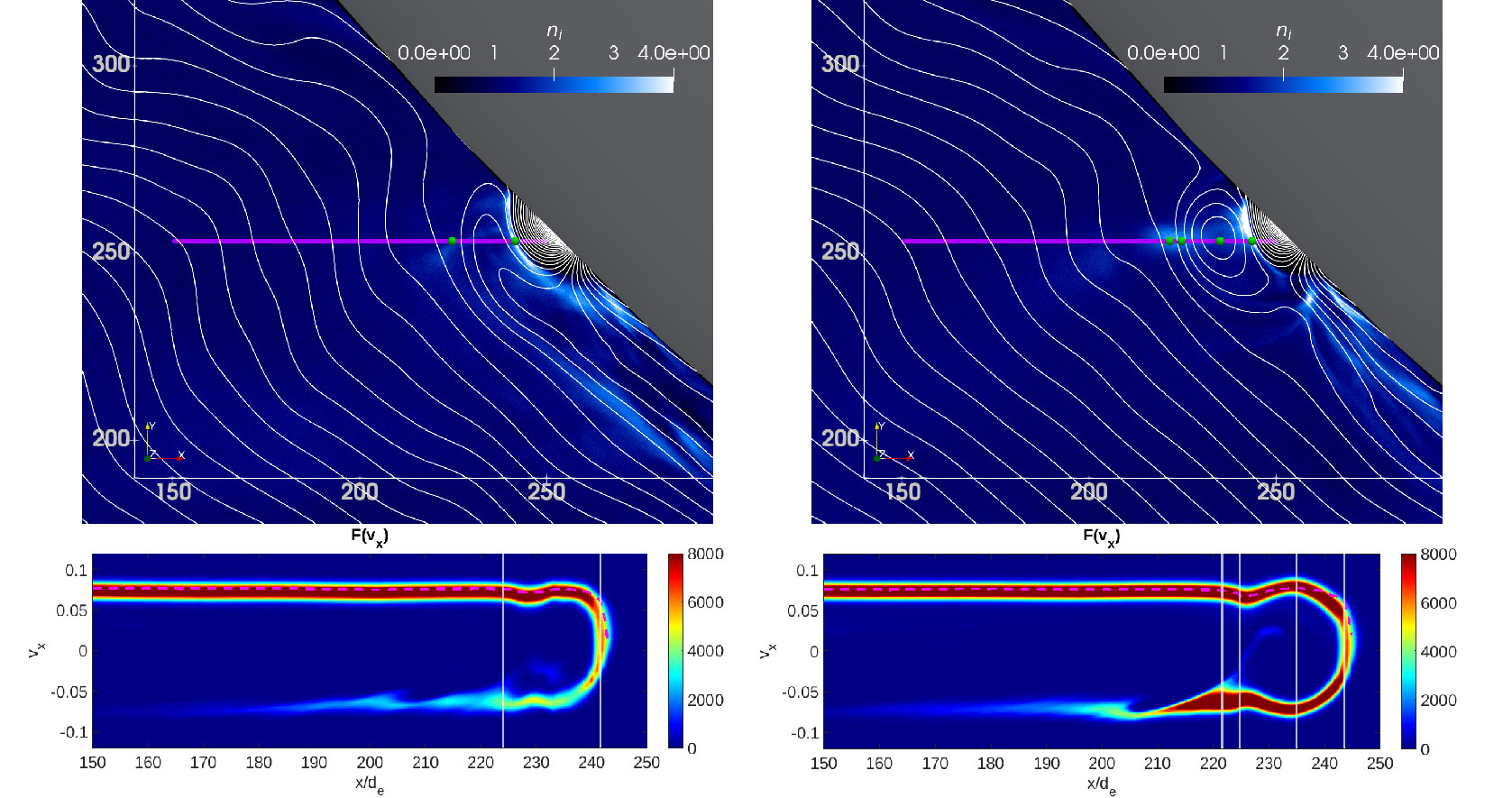}
    \caption{Top: Contours show magnetic field-lines and the color scale shows ion density. Bottom: Ion velocity distribution functions $F(v_x)$ computed along the horizontal line. The white vertical lines correspond to the green dots in the top plots. The magenta lines show the predicted $v_x$ for $v_x(t=0)=v_{\textrm{SW}}$ which is accelerated/decelerated by an electric field $E_x = -\partial_x B_y^2/(2\mu_0n_ee)$. The left column is without an magnetic island, and the right column has an island present with radius $\sim 0.25 d_i$. }
    \label{fig:spectra}
\end{figure}

Figure~\ref{fig:spectra} shows a comparison of the reduced ion distribution function, $f(x,v_x),$ at two times - without (left) and with (right) a magnetic island present in the interaction region. In both cases, the incoming solar wind population with mean $\overline{v}_x = v_{SW}$ experiences significant reflection at the boundary of the closed LCMA field-lines. However, the fraction of ions reflected is significantly more when the magnetic island is present. The large number of reflected ions gives a visible `spike' feature in the plasma density (top right), which is locally increased to $\approx 3 \times$ the background value. By integrating inflowing and reflected parts of the distribution function over the $y$-extent of the LCMA dipole field, the total fraction of ions reflected at the location of the density spike feature is $40\%$ without the island, and increasing to $70\%$ in the case where the island is present. 

At both times, with or without the island present, we find that the ions are reflected by Hall electric fields resulting from the different magnetizations of the ions and electrons. The electrons, being magnetized, are frozen into the magnetic field and deflected by the magnetic barrier, while the ions having a large gyro-radius do not experience significant magnetic deflection. However, as described below, the resulting strong Hall electric fields can slow down and reflect the ions.

Consider an ion with mass $m_i$ and charge $q_i$ moving with the mean solar wind velocity $v_{x0} = v_{SW}$. If it is assumed that the ion is sufficiently unmagnetized, the velocity change in response to an electric field is given by
\begin{equation}\label{ionslowingdown}v_{x} \approx v_{x0} + (q_i/m_i) \int{E_x dt}.\end{equation} 
Now, assuming that the electron fluid obeys the frozen-in condition, the Hall electric field has dominant term given by
\begin{equation}\label{efield}E_x \approx -\frac{j_{z}B_y}{en_e} \approx -\frac{\partial_x B_y^2}{2n_e \mu_0 e}.\end{equation} 
(This expression can also be derived as $\boldsymbol{E} = -\boldsymbol{\nabla}\phi$ using the electrostatic potential $\phi$ derived in~\cite{bamford2012prl,cruz2017pop}).

The magenta dashed line plotted in the bottom panels of Fig.~\ref{fig:spectra} is found by integrating Eq.~(\ref{ionslowingdown}) with the Hall electric field from Eq.~(\ref{efield}), and closely matches the shape of the distribution functions. Thus, the Hall electric field is responsible for the decelerations and accelerations of ions in the calculated distribution function. For the snapshot with the magnetic island, the IMF magnetic field first piles-up to increase $B_y^2$ on the left edge of the magnetic island, passes through zero at the island center, before again piling up at the LCMA closed field boundary. This is the cause of the small dip in the inflowing ion velocities at $x= 228 d_e$, followed by the subsequent acceleration up to $x=235 d_e$ and final deceleration/reflection features. Even though there is no magnetic island in the left hand plots, the displacement of the X-point up towards the Moon's surface gives similar qualitative magnetic field-line profiles along the magenta horizontal line where the distribution function is calculated. The presence of the magnetic island however gives stronger increase of $B_y^2$ due to pile-up, thus giving stronger deceleration/acceleration features, and reflecting a larger fraction of ions back upstream.

\begin{figure}
    \centering
    \includegraphics[width=0.95\textwidth]{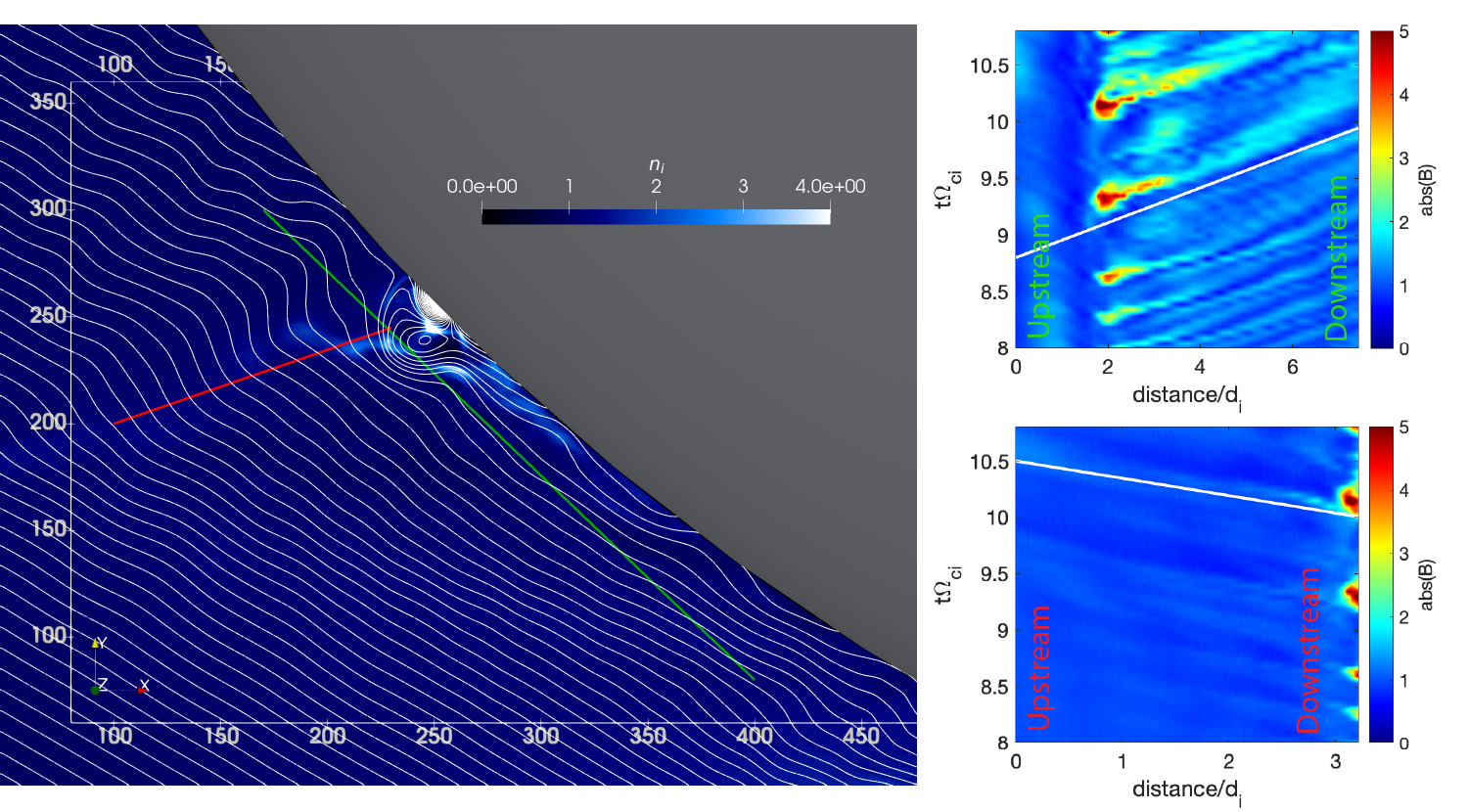}
    \caption{Left: Ion density (color scale) and magnetic field-lines (white) at time $t\Omega_{ci}=10.15$ (when there is a magnetic island present). The spatial coordinates are given in electron skin depth ($1\, d_e \approx 1.5$ km). Right: Absolute value of the magnetic field ($|\boldsymbol{B}|/B_0$) sampled from the green line (top right plot) and red line (bottom right plot), plotted against distance along each line from upstream to downstream (horizontal axes) in ion-skin depths ($1\,d_i = \sqrt{1836}\,d_e$), and time (vertical axes) in $\Omega_{ci}^{-1}$. The white lines indicate $v=v_{SW}$ (top) and $v=-v_{SW}$ (bottom).}
    \label{fig:perturbations}
\end{figure}

Figure~\ref{fig:perturbations} shows how fluctutations in density (left plot) and magnetic field strength (right plots), caused by the quasi-periodic formation of islands, propagates downstream above the lunar surface (green line, top right plot), and back upstream into the solar wind (red line, bottom right plot). A density spike feature caused by reflected ions (see above) is visible next to the red line in the left hand plot. The magnetic compressions are as large as $\boldsymbol{B}/B_{IMF} \approx 5$ at the island but reduce to $\boldsymbol{B}/B_{IMF} \approx 1.7$ by $\approx 2 d_i$ downstream. The perturbations are carried with the solar wind velocity $v_{SW}$ downstream (white line). 
As shown in the bottom right panel, such magnetic compressions might also be detected a short distance upstream of the LCMA fields, even in the direction perpendicular to the inflowing IMF (red line), but these signatures drop off to small values over a shorter distance ($\lesssim 15\%$ by $\approx 0.5d_i$ upstream). The white line shows a propagation velocity $v=-v_{SW}$, suggesting that the disturbances propagate back upstream by the reflected ion population. Note that for the simulation parameters specified in Sec.~\ref{sec:params}, the characteristic velocities are well separated with 
\begin{equation*}\frac{v_{th,e}}{v_{SW}} (= 3.6) \gg 1 \gg \frac{v_{A}}{v_{SW}} (= 0.15) \gg \frac{v_{th,i}}{v_{SW}} (= 0.084).\end{equation*}

\subsection{Discussion}

The simulations described indicate the occurrence of a bursty electron only reconnection due to the repeated formation and ejection of magnetic islands, and the generation of strong electron parallel flows along the separator lines of the X-points in the SW-LCMA interation region. Moreover, this system exhibits interesting dynamical behavior that is not seen in more typically studied magnetospheric systems with more separation between the shock and reconnection physics. Namely, the quasi-periodic formation of islands leads to a time variabilty of the solar wind stand-off and pressure balance, and gives sharp modulations the proton reflectivity via changes in the Hall electric field. These protons in turn drive increased fluctuations downstream and upstream in the SW to generate observable magnetic field compressions. 

Our simulations use a realistic mass-ratio and are the first to show signatures of electron-only reconnection and the formation of magnetic islands at sub-ion-scale crustal fields. For example, \citep{deca2014prl} report having observed no signatures of magnetic reconnection from their simulations. While our simulations also do not show the ion jets that are characteristic of macroscopic reconnection, the formation of islands and accelerated parallel electron flows along the X-point separator lines indicate electron-only reconnection. There are several differences to our simulations that could explain this discrepancy. Firstly, at lower values of the ion-to-electron mass ratio $m_i/m_e < 400$, we have observed that the formation of islands becomes more marginal, and the SW-LCMA interaction region becomes stable for lower values of $m_i/m_e \approx 100$. Although a formal boundary layer analysis of the tearing mode is difficult for the LCMA current sheet, due to the complex geometry of the fields and flows in the region and the lack of scale separation that makes the calculation tractable, some insight can be gained by modeling it as a Harris sheet with electron-scale thickness. By neglecting additional stabilization mechanisms such as shear flows, a lower bound for the instability threshold can be estimated when the parameter $\Delta^\prime > 0$, where
\begin{equation}\Delta^\prime = \frac{2}{\delta} \left(\frac{1}{k\delta} - k\delta\right).\end{equation}

The longest wavelength mode corresponding to a single magnetic island has $k=k_{\textrm{min}} = 2\pi/L$ where $L$ is the length of the current sheet. Thus $\Delta^\prime > 0$ for $L > 2\pi \delta$. In the time between the formation of magnetic islands, we measure the FWHM thickness $\delta=1.2 d_e$ and aspect ratio $L/\delta \approx 12.5$, thus exceeding the $\Delta^\prime$-threshold for a tearing instability. On the other hand, for a simulation with $m_i/m_e=100$ and the same resolution ($\Delta x/d_e = 0.06$), we find $\delta\approx 1.3 d_e$, but the length decreases to $6.2 d_e$ (as the length of the current distribution is limited by spatial-scale of the LCMA fields that is set in terms of ion inertial lengths). This gives an aspect ratio $L/\delta = 4.8$ (below the threshold) that fits with the absence of islands forming in that simulation. This may explain why islands have not been seen in other kinetic simulations of sub-ion-gyroradius scale LCMA run with lower values of $m_i/m_e$. We note that for mini-magnetospheres with larger than ion-scale stand-off distances, magnetic islands have been observed in PIC simulations~\citep{cruz2017pop}. 

The quasi-neutral hybrid simulations of~\cite{giacalone2015hybrid} use an IMF angle such that a magnetic X-point forms to one side of the main LCMA field. They find that strong magnetic gradients between the X-type neutral point and the stronger LCMA surface field can generate an additional spatially separated Hall electric field, and this is found to give rise to a distinct secondary population of reflected protons. Our study supports the observation that the X and O-type magnetic topological points can modify the Hall electric structure and modulate the reflected proton flux. However, some of the main features reported in this letter regarding the time dependence of the reflected proton flux, the magnetic island formation, and the accelerated electron flows along the magnetic field-lines were not observed in the kinetic-ion fluid-electron hybrid simulations reported by~\cite{giacalone2015hybrid}, presumably because these processes depend on electron kinetic physics that is not present in the hybrid model~\citep{winske2023hybrid}. 

The use of two-dimensional simulations allows us to simulate larger spatial domains ($12.5\times 12.5d_i$, $8192\times 8192$ cells) and mass-ratios ($m_i/m_e=1836$) than would be feasible at present in 3D domains. However, this does impose a key limitation: the effectively infinite extent of a line dipole source in $Z$ in our 2D simulations constrains the electrons to flow around the LCMA in the $X$-$Y$ plane only, whereas in 3D the strong electron flow in the $Z$ direction would be limited in length by the extent of the LCMA structure in that dimension. As the magnetic field becomes frozen into the electron flows on sub-ion scales, it can experience different advection by such flows, which can produce a different distribution of current about the LCMA in 3D vs 2D. Future work will explore 3D effects on this process at similar mass ratios to those used in our 2D study, and also study the influence of more complex multiple-dipole models of the LCMA fields~\citep[as in e.g.][]{harnett2003,xie2015hall}, to give a larger length of the SW-LCMA interaction region that may increase the likelihood of flux-rope formation due to tearing instabilties.

It may be possible to observe the quasi-periodic propagating magnetic field compressions and density spikes shown in Fig. 4 with the ARTEMIS spacecraft at periapsis, as they can be distinguished from fluctuations in the background solar wind through pitch-angle distributions. There may also a possibility to experimentally study the formation of magnetic islands in kinetic scale magnetic dipole configurations that resemble mini-magnetospheres, which can be formed inside laser driven plasmas~\citep{schaeffer2022}. Indeed, small-scale magnetic islands are becoming quite commonly measured in electron-scale current sheets in other laboratory reconnection experiments~\citep{olson2016islands,jara2016,kamio2018}, including in the electron-only regime~\citep{olson2016islands}.

\begin{acknowledgments}
This research was supported by the NASA MMS and HERMES missions, and by the DOE Frontier Plasma Science Program. J.H. and R. S. were supported by the THEMIS-ARTEMIS mission (NASA contract NAS5-02099), and the Solar System Exploration Research Virtual Institute (80NSSC20M0022).
\end{acknowledgments}
%

\vspace{5mm}


\begin{thebibliography}{}
\expandafter\ifx\csname natexlab\endcsname\relax\def\natexlab#1{#1}\fi
\providecommand{\url}[1]{\href{#1}{#1}}
\providecommand{\dodoi}[1]{doi:~\href{http://doi.org/#1}{\nolinkurl{#1}}}
\providecommand{\doeprint}[1]{\href{http://ascl.net/#1}{\nolinkurl{http://ascl.net/#1}}}
\providecommand{\doarXiv}[1]{\href{https://arxiv.org/abs/#1}{\nolinkurl{https://arxiv.org/abs/#1}}}

\bibitem[{Bamford {et~al.}(2012)Bamford, Kellett, Bradford, Norberg, Thornton,
  Gibson, Crawford, Silva, Gargat{\'e}, \& Bingham}]{bamford2012prl}
Bamford, R., Kellett, B., Bradford, W., {et~al.} 2012, Physical Review Letters,
  109, 081101

\bibitem[{Bamford {et~al.}(2016)Bamford, Alves, Cruz, Kellett, Fonseca, Silva,
  Trines, Halekas, Kramer, Harnett, {et~al.}}]{bamford2016}
Bamford, R., Alves, E., Cruz, F., {et~al.} 2016, The Astrophysical Journal,
  830, 146

\bibitem[{Bird {et~al.}(2021)Bird, Tan, Luedtke, Harrell, Taufer, \&
  Albright}]{bird2021vpic}
Bird, R., Tan, N., Luedtke, S.~V., {et~al.} 2021, IEEE Transactions on Parallel
  and Distributed Systems, 33, 952

\bibitem[{Birdsall \& Langdon(2004)}]{birdsall2004}
Birdsall, C.~K., \& Langdon, A.~B. 2004, Plasma physics via computer simulation
  (CRC press)

\bibitem[{{Bowers} {et~al.}(2008){Bowers}, {Albright}, {Yin}, {Bergen}, \&
  {Kwan}}]{bowers08}
{Bowers}, K.~J., {Albright}, B.~J., {Yin}, L., {Bergen}, B., \& {Kwan},
  T.~J.~T. 2008, Phys. Plasmas, 15, 055703, \dodoi{10.1063/1.2840133}

\bibitem[{Colburn {et~al.}(1967)Colburn, Currie, Mihalov, \&
  Sonett}]{colburn1967}
Colburn, D.~S., Currie, R.~G., Mihalov, J.~D., \& Sonett, C.~P. 1967, Science,
  158, 1040, \dodoi{10.1126/science.158.3804.1040}

\bibitem[{Coleman {et~al.}(1972)Coleman, Schubert, Russell, \&
  Sharp}]{coleman1972satellite}
Coleman, P.~J., Schubert, G., Russell, C., \& Sharp, L. 1972, The moon, 4, 419

\bibitem[{Cruz {et~al.}(2017)Cruz, Alves, Bamford, Bingham, Fonseca, \&
  Silva}]{cruz2017pop}
Cruz, F., Alves, E., Bamford, R., {et~al.} 2017, Physics of Plasmas, 24

\bibitem[{Deca {et~al.}(2014)Deca, Divin, Lapenta, Lemb{\`e}ge, Markidis, \&
  Hor{\'a}nyi}]{deca2014prl}
Deca, J., Divin, A., Lapenta, G., {et~al.} 2014, Physical review letters, 112,
  151102

\bibitem[{Deca {et~al.}(2015)Deca, Divin, Lemb{\`e}ge, Hor{\'a}nyi, Markidis,
  \& Lapenta}]{deca2015}
Deca, J., Divin, A., Lemb{\`e}ge, B., {et~al.} 2015, Journal of Geophysical
  Research: Space Physics, 120, 6443

\bibitem[{Deca {et~al.}(2018)Deca, Divin, Lue, Ahmadi, \&
  Hor{\'a}nyi}]{deca2018reiner}
Deca, J., Divin, A., Lue, C., Ahmadi, T., \& Hor{\'a}nyi, M. 2018,
  Communications Physics, 1, 12

\bibitem[{Dyal {et~al.}(1974)Dyal, Parkin, \& Daily}]{dyal1974}
Dyal, P., Parkin, C.~W., \& Daily, W.~D. 1974, Reviews of Geophysics, 12, 568,
  \dodoi{https://doi.org/10.1029/RG012i004p00568}

\bibitem[{Eastwood {et~al.}(2008)Eastwood, Brain, Halekas, Drake, Phan,
  {\O}ieroset, Mitchell, Lin, \& Acu{\~n}a}]{eastwood2008mars}
Eastwood, J., Brain, D., Halekas, J., {et~al.} 2008, Geophysical Research
  Letters, 35

\bibitem[{Fatemi {et~al.}(2014)Fatemi, Holmstr{\"o}m, Futaana, Lue, Collier,
  Barabash, \& Stenberg}]{fatemi2014hybrid}
Fatemi, S., Holmstr{\"o}m, M., Futaana, Y., {et~al.} 2014, Journal of
  Geophysical Research: Space Physics, 119, 6095

\bibitem[{Fatemi {et~al.}(2013)Fatemi, Holmström, Futaana, Barabash, \&
  Lue}]{fatemi2013}
Fatemi, S., Holmström, M., Futaana, Y., Barabash, S., \& Lue, C. 2013,
  Geophysical Research Letters, 40, 17,
  \dodoi{https://doi.org/10.1029/2012GL054635}

\bibitem[{Futaana {et~al.}(2003)Futaana, Machida, Saito, Matsuoka, \&
  Hayakawa}]{futaana2003}
Futaana, Y., Machida, S., Saito, Y., Matsuoka, A., \& Hayakawa, H. 2003,
  Journal of Geophysical Research: Space Physics, 108, SMP

\bibitem[{Giacalone \& Hood(2015)}]{giacalone2015hybrid}
Giacalone, J., \& Hood, L. 2015, Journal of Geophysical Research: Space
  Physics, 120, 4081

\bibitem[{Halekas {et~al.}(2009)Halekas, Eastwood, Brain, Phan, {\O}ieroset, \&
  Lin}]{halekas2009mars}
Halekas, J., Eastwood, J., Brain, D., {et~al.} 2009, Journal of Geophysical
  Research: Space Physics, 114

\bibitem[{Halekas {et~al.}(2017)Halekas, Poppe, Lue, Farrell, \&
  McFadden}]{halekas2017}
Halekas, J., Poppe, A., Lue, C., Farrell, W., \& McFadden, J. 2017, Journal of
  Geophysical Research: Space Physics, 122, 6240

\bibitem[{Halekas {et~al.}(2014)Halekas, Poppe, McFadden, Angelopoulos,
  Glassmeier, \& Brain}]{halekas2014}
Halekas, J., Poppe, A., McFadden, J., {et~al.} 2014, Geophysical Research
  Letters, 41, 7436

\bibitem[{Harada {et~al.}(2018)Harada, Halekas, DiBraccio, Xu, Espley,
  Mcfadden, Mitchell, Mazelle, Brain, Hara, {et~al.}}]{harada2018mars}
Harada, Y., Halekas, J., DiBraccio, G., {et~al.} 2018, Geophysical Research
  Letters, 45, 4550

\bibitem[{Harnett \& Winglee(2003)}]{harnett2003}
Harnett, E.~M., \& Winglee, R.~M. 2003, Journal of Geophysical Research: Space
  Physics, 108

\bibitem[{Hood \& Schubert(1980)}]{hood1980}
Hood, L.~L., \& Schubert, G. 1980, Science, 208, 49,
  \dodoi{10.1126/science.208.4439.49}

\bibitem[{Jara-Almonte {et~al.}(2016)Jara-Almonte, Ji, Yamada, Yoo, \&
  Fox}]{jara2016}
Jara-Almonte, J., Ji, H., Yamada, M., Yoo, J., \& Fox, W. 2016, Physical Review
  Letters, 117, 095001

\bibitem[{Ji {et~al.}(2022)Ji, Daughton, Jara-Almonte, Le, Stanier, \&
  Yoo}]{ji2022natrevphys}
Ji, H., Daughton, W., Jara-Almonte, J., {et~al.} 2022, Nature Reviews Physics,
  4, 263

\bibitem[{Kamio {et~al.}(2018)Kamio, Inomoto, Yamasaki, Yamada, Cheng, \&
  Ono}]{kamio2018}
Kamio, S., Inomoto, M., Yamasaki, K., {et~al.} 2018, Physics of Plasmas, 25

\bibitem[{Lin {et~al.}(1998)Lin, Mitchell, Curtis, Anderson, Carlson, McFadden,
  Acuna, Hood, \& Binder}]{lin1998}
Lin, R., Mitchell, D., Curtis, D., {et~al.} 1998, Science, 281, 1480

\bibitem[{Lue {et~al.}(2011)Lue, Futaana, Barabash, Wieser, Holmström,
  Bhardwaj, Dhanya, \& Wurz}]{lue2011}
Lue, C., Futaana, Y., Barabash, S., {et~al.} 2011, Geophysical Research
  Letters, 38, \dodoi{https://doi.org/10.1029/2010GL046215}

\bibitem[{Lyon {et~al.}(1967)Lyon, Bridge, \& Binsack}]{lyon1967explorer}
Lyon, E.~F., Bridge, H.~S., \& Binsack, J.~H. 1967, Journal of Geophysical
  Research, 72, 6113

\bibitem[{Nishino {et~al.}(2009)Nishino, Fujimoto, Maezawa, Saito, Yokota,
  Asamura, Tanaka, Tsunakawa, Matsushima, Takahashi, {et~al.}}]{nishino2009}
Nishino, M., Fujimoto, M., Maezawa, K., {et~al.} 2009, Geophysical Research
  Letters, 36

\bibitem[{Olson {et~al.}(2016)Olson, Egedal, Greess, Myers, Clark, Endrizzi,
  Flanagan, Milhone, Peterson, Wallace, {et~al.}}]{olson2016islands}
Olson, J., Egedal, J., Greess, S., {et~al.} 2016, Physical review letters, 116,
  255001

\bibitem[{Owen {et~al.}(1996)Owen, Lepping, Ogilvie, Slavin, Farrell, \&
  Byrnes}]{owen1996}
Owen, C.~J., Lepping, R.~P., Ogilvie, K.~W., {et~al.} 1996, Geophysical
  Research Letters, 23, 1263, \dodoi{https://doi.org/10.1029/96GL01354}

\bibitem[{Phan {et~al.}(2018)Phan, Eastwood, Shay, Drake, Sonnerup, Fujimoto,
  Cassak, {\O}ieroset, Burch, Torbert, {et~al.}}]{phan2018electron}
Phan, T., Eastwood, J.~P., Shay, M., {et~al.} 2018, Nature, 557, 202

\bibitem[{Purucker \& Nicholas(2010)}]{purucker2010global}
Purucker, M.~E., \& Nicholas, J.~B. 2010, Journal of Geophysical Research:
  Planets, 115

\bibitem[{Russell \& Lichtenstein(1975)}]{russell1975compressions}
Russell, C., \& Lichtenstein, B. 1975, Journal of Geophysical Research, 80,
  4700

\bibitem[{Sawyer {et~al.}(2023)Sawyer, Halekas, Bonnell, Chen, McFadden,
  Glassmeier, Harada, \& Stanier}]{sawyer2023}
Sawyer, R.~P., Halekas, J.~S., Bonnell, J.~W., {et~al.} 2023, Geophysical
  Research Letters, 50, e2023GL104733,
  \dodoi{https://doi.org/10.1029/2023GL104733}

\bibitem[{Schaeffer {et~al.}(2022)Schaeffer, Cruz, Dorst, Cruz, Heuer,
  Constantin, Pribyl, Niemann, Silva, \& Bhattacharjee}]{schaeffer2022}
Schaeffer, D.~B., Cruz, F.~D., Dorst, R.~S., {et~al.} 2022, Physics of Plasmas,
  29, 042901, \dodoi{10.1063/5.0084353}

\bibitem[{Winske {et~al.}(2023)Winske, Karimabadi, Le, Omidi, Roytershteyn, \&
  Stanier}]{winske2023hybrid}
Winske, D., Karimabadi, H., Le, A.~Y., {et~al.} 2023, in Space and
  Astrophysical Plasma Simulation: Methods, Algorithms, and Applications
  (Springer), 63--91

\bibitem[{Xie {et~al.}(2015)Xie, Li, Zhang, Feng, Wang, Zhang, \&
  Kong}]{xie2015hall}
Xie, L., Li, L., Zhang, Y., {et~al.} 2015, Journal of Geophysical Research:
  Space Physics, 120, 6559

\bibitem[{Xu {et~al.}(2020)Xu, Poppe, Halekas, \& Harada}]{xu2020reflected}
Xu, S., Poppe, A.~R., Halekas, J.~S., \& Harada, Y. 2020, Journal of
  Geophysical Research: Space Physics, 125, e2020JA028154

\bibitem[{Yee(1966)}]{yee1966}
Yee, K. 1966, IEEE Transactions on antennas and propagation, 14, 302

\end{thebibliography}
\end{document}